\documentstyle[12pt]{article}
\def\figcap{\section*{Figure Captions\markboth
        {FIGURECAPTIONS}{FIGURECAPTIONS}}\list
        {Figure \arabic{enumi}:\hfill}{\settowidth\labelwidth{Figure
999:}
        \leftmargin\labelwidth
        \advance\leftmargin\labelsep\usecounter{enumi}}}
 \relax
\def\ap#1#2#3{Ann.\ Phys.\ (NY) #1 (19#3) #2}

\def\jmp#1#2#3{J.\ Math.\ Phys.\ #1 (19#3) #2}

\def\np#1#2#3{Nucl.\ Phys.\ B#1 (19#3) #2}
\def\pl#1#2#3{Phys.\ Lett.\ #1B (19#3) #2}
\def\pr#1#2#3{Phys.\ Rev.\ D #1 (19#3) #2}
\def\prb#1#2#3{Phys.\ Rev.\ B #1 (19#3) #2}
\def\prep#1#2#3{Phys.\ Rep.\ #1 (19#3) #2}

\def\rmp#1#2#3{Rev.\ Mod.\ Phys.\ #1 (19#3) #2}

\def\zp#1#2#3{Zeit.\ Phys.\ C#1 (19#3) #2}
\def\cmp#1#2#3{Comm.\ Math.\ Phys.\ #1 (19#3) #2}
\def\cmp#1#2#3{Comm.\ Math.\ Phys.\ #1 (19#3) #2}
\def\nc#1#2#3{Il Nuovo Cimento #1A (19#3) #2}
\def\lnc#1#2#3{Lettere al Nuovo Cimento #1 (19#3) #2}
\def\apa#1#2#3{Acta Phy. Austriaca #1 (19#3) #2}
\newcounter{hran}

\def\bmini#1{\setcounter{hran}{\value{equation}}
\refstepcounter{hran} \setcounter{equation}{#1}
\renewcommand{\theequation}{\thehran\alph{equation}}
              \begin{eqnarray}  }

\def\bminiG#1{
          \setcounter{hran}{\value{equation}}
          \refstepcounter{hran}
          \setcounter{equation}{-1}
          \renewcommand{\theequation}{\thehran\alph{equation}}
          \refstepcounter{equation}
    \label{#1}
          \begin{eqnarray}          }

\def\emini{\end{eqnarray}\setcounter{equation}{\value{hran}}
\renewcommand{\theequation}{\arabic{equation}}}
\newskip\humongous \humongous=0pt plus 1000pt minus 1000pt
\def\caja{\mathsurround=0pt} \def\eqalign#1{\,\vcenter{\openup1\jot
\caja   \ialign{\strut \hfil$\displaystyle{##}$&$
\displaystyle{{}##}$\hfil\crcr#1\crcr}}\,} \newif\ifdtup

\def\frac#1#2{ {{#1} \over {#2} }}
\def\ie{\hbox{\it i.e.}{ }}      
\def\eg{\hbox{\it e.g.}{ }}      
\def\half{\mbox{\small $\frac{1}{2}$}}
\def\partder#1{{\partial   \over\partial #1}}

\def\re#1{(\ref{#1})}
\def\beq{\begin{equation}}
\def\eeq{\end{equation}}
\def\beeq{\begin{eqnarray}}
\def\beeqn{\begin{eqnarray*}}
\def\eeeq{\end{eqnarray}}
\def\eeeqn{\end{eqnarray*}}
\def\se{S_{\mbox{\footnotesize{eff}}}}
\def\ser{S_{\mbox{\footnotesize{eff,rel}}}}
\def\sei{S_{\mbox{\footnotesize{eff,irr}}}}
\def\set{S_{\mbox{\scriptsize{eff}}}}
\def\si{S_{\mbox{\footnotesize{int}}}}
\def\Gr{\G_{\mbox{\footnotesize{rel}}}}
\def\Grt{\tilde \G_{\mbox{\footnotesize{rel}}}}
\def\Gi{\G_{\mbox{\footnotesize{irr}}}}
\def\Dr{\D_{\mbox{\footnotesize{rel}}}}
\def\Di{\D_{\mbox{\footnotesize{irr}}}}
\def\DGr{\D_{\G,\mbox{\footnotesize{rel}}}}

\def\bchi{\bar \chi}

\def\bc{\bar c}
\def\bp{\bar p}

\def\r{\rho}
\def\br{\bar \rho}
\def\de{\delta}
\def\s{\sigma}

\def\S{\Sigma}
\def\G{\Gamma}
\def\eps{\epsilon}

\def\L{\Lambda}
\def\l{\lambda}
\def\g{\gamma}
\def\D{\Delta}
\def\d4#1{\frac {d^4 {#1} }{(2\pi)^4}}
\def\dL{\L \partial_\L }

\def\UV{$\L_0\to\infty\;$}
\def\IR{$\L\to 0\;$}
\def\bit{\begin{itemize}}
\def\eit{\end{itemize}}
\def\ben{\begin{enumerate}}
\def\een{\end{enumerate}}

\def\nome#1{{\label{#1}}}

\begin{document}
\begin{titlepage}
\renewcommand{\thefootnote}{\fnsymbol{footnote}}
\begin{flushright}
     UPRF 94-412\\
     October 1994 \\
\end{flushright}
\par \vskip 10mm
\begin{center}
{\Large \bf
BRS symmetry for
Yang-Mills theory\\
with exact renormalization group
\footnote{Research supported in part by MURST, Italy}}
\end{center}
\par \vskip 2mm
\begin{center}
        {\bf M.\ Bonini, M.\ D'Attanasio} \\
        Dipartimento di Fisica, Universit\`a di Parma and\\
        INFN, Gruppo Collegato di Parma, Italy\\
        and\\
        {\bf G.\ Marchesini}\\
        Dipartimento di Fisica, Universit\`a di Milano and\\
        INFN, Sezione di Milano, Italy
\end{center}
\par \vskip 2mm
\begin{center} {\large \bf Abstract} \end{center}
\begin{quote}
In the exact renormalization group (RG) flow in the infrared cutoff
$\L$ one needs boundary conditions.
In a previous paper on $SU(2)$ Yang-Mills theory we proposed to use
the nine physical relevant couplings of the effective action
as boundary conditions at the physical point $\L=0$
(these couplings are defined at some non-vanishing subtraction
point $\mu \ne 0$).
In this paper we show perturbatively that it is possible to
appropriately fix these couplings in such a way that the full set of
Slavnov-Taylor (ST) identities are satisfied.
Three couplings are given by the vector and ghost wave function
normalization and the three vector coupling at the subtraction point;
three of the remaining six are vanishing (\eg the vector mass)
and the others are expressed by irrelevant vertices
evaluated at the subtraction point.
We follow the method used by Becchi to prove ST identities in the
RG framework. There the boundary conditions are given at a non-physical
point $\L=\L' \ne 0$, so that one avoids the need of a non-vanishing
subtraction point.
\end{quote}
\end{titlepage}

\section{Introduction}
Renormalization group (RG) formulation \cite{W} provides the
most physical way to deal with the ultraviolet (UV) singularities.
Following a suggestion of Polchinski \cite{P} (see also
\cite{G}-\cite{B})
the exact RG formulation has been used recently to give self-contained
and simple perturbative rederivations of many properties such as
renormalizability \cite{P},\cite{B}-\cite{ren}, infrared finiteness of
massless theories \cite{BDM,IR}, operator product expansion \cite{OPE},
decoupling theorem \cite{dec}.
Some approximations have also been attempted within this context \cite{approx}.

Its application has been recently extended to gauge and
chiral gauge theories \cite{B},\cite{YM}-\cite{chi}.
The challenging problem here is the general fact that local
gauge symmetry typically conflicts with the presence of a
momentum cutoff. For the $SU(2)$ Yang-Mills theory it is shown in
ref.~\cite{B} and in the present paper that,
in spite of the explicit breaking of gauge symmetry,
the Slavnov-Taylor (ST) identities can be implemented perturbatively
by appropriately fixing the boundary conditions on the RG flow.
Thus the RG formulation provides an alternative
to dimensional regularization \cite{dim} of gauge theories.

To describe the situation we briefly recall the exact RG formulation
for gauge theories.
Consider the Wilsonian effective action $\se[\phi,\g;\L]$
with $\phi$ and $\g$ the fields and the BRS sources and $\L$
an infrared (IR) cutoff. This functional is obtained by path
integration over the fields with frequencies
above $\L$ and up to an UV cutoff $\L_0\,$.
At $\L=0$ and \UV one has performed the full path integral and
therefore from the functional $\se[\phi,\g;\L=0]$ we can obtain
the physical Green functions or the one particle irreducible vertices.
Actually the vertices of $\se[\phi,\g;\L=0]$ are the amputated
connected Green functions. We never write in the functionals
the UV cutoff $\L_0$ since we always understand \UV, which is possible
due to perturbative renormalizability.
The functional $\se[\phi,\g;\L]$ satisfies
an evolution equation in the IR cutoff,
which is obtained by observing that the path integral over
all frequencies can be done first by integrating
over the frequencies above $\L$ and then below $\L$. The fact that
the result does not depend on $\L$ gives the mentioned evolution
equation.

The exact RG equation is non-perturbative but it can be solved
perturbatively once the boundary conditions are given at some value of
$\L$.
An important point concerning the boundary conditions is the
distinction between relevant and irrelevant parts of the functional
$\se[\phi,\g;\L]$ (and of any other functional such as the
physical effective action $\G[\phi,\g]$).
As usual the vertices with negative mass dimension are called
``irrelevant'' and contribute to $\sei[\phi,\g;\L]$.
The remaining part $\ser[\phi,\g;\L]$ contains only a finite
number of parameters with non-negative dimension (relevant
parameters) which, for $\L \ne 0$, can be defined for instance as
the first coefficients of the Taylor expansion of $\se[\phi,\g;\L]$
around vanishing momenta.

The UV value $\L=\L_0$ is the obvious value where to fix the boundary
conditions for $\sei[\phi,\g;\L]$.
For \UV one expects that all vertices with negative dimension vanish
thus one sets $\sei[\phi,\g;\L=\L_0] \to 0 $ for \UV.

Finally the key problem is to fix the boundary conditions for
the finite number of relevant parameters in $\ser[\phi,\g;\L]$.
Here is where one implements in the theory both the physical
parameters (such as masses, couplings, wave function normalizations)
and the symmetry, \ie ST identities.

In ref.~\cite{YM} we considered the $SU(2)$ Yang-Mills theory which
has nine relevant parameters.
We proposed to fix these nine parameters at the physical point $\L=0$.
Since the theory involves zero mass fields, at $\L=0$ one should define
these parameters as the values of some vertices at non-vanishing
subtraction points.
The reason for selecting $\L=0$ is that $\se[\phi,\g;\L=0]$ is
related via Legendre transform to the effective action $\G[\phi,\g]$.
Therefore some of these parameters are fixed at the physical values
of the couplings, masses and wave function normalizations.
We proposed to obtain the remaining parameters
by imposing some of the ST identities
and showed that this procedure can be implemented perturbatively.
Therefore by construction some of the ST identities are satisfied,
thus the fundamental question is whether the effective action
$\G[\phi,\g]$ computed by this procedure does indeed satisfy the
full set of ST identities.
In the paper of ref.~\cite{YM} we were able to prove this only partially
at one loop level.

Recently Becchi \cite{B} considered the exact RG formulation of
the $SU(2)$ Yang-Mills theory and imposed the boundary conditions for
the relevant parameters in $\ser[\phi,\g;\L]$ at a non-physical
point $\L=\L'\ne 0$ so that the relevant parameters can be defined
by expanding the vertices around vanishing momenta.
In this way the relevant parameters are not directly
related to physical couplings in the effective action $\G[\phi,\g]$
but the analysis of relevant parts of ST identities becomes easy.
By using algebraic methods he was able to prove that the full
set of ST identities can indeed be satisfied perturbatively.
He identified a functional $\Delta_{\mbox{\scriptsize{eff}}}[\phi,\g;\L]$
which gives the ``defect'' to ST identities and then showed
that one can obtain perturbatively the nine relevant
parameters in $\ser[\phi,\g;\L]$
by solving the fine tuning equation
$$
\Delta_{\mbox{\scriptsize{eff,rel}}}[\phi,\g;\L]=0
$$
for the relevant part of the ST defect.

In the present paper we apply the same method to the case in which the
boundary conditions are given at the physical point $\L=0$.
In this case the operation of extracting the relevant parameters become
rather complex since one has to use non-vanishing subtraction points.
We are able to generalize the Becchi's proof to this case and we
explicitly prove that the procedure suggested in our previous paper
gives indeed an effective action which satisfies perturbatively ST
identities.
Moreover we explicitly express the solution of the fine tuning equation
by giving in terms of physical vertices those relevant parameters
which are not fixed by the physical couplings and wave
function normalizations.

This paper is organized as follows.
In section $2$ we recall the RG formulation for the $SU(2)$ Yang-Mills
theory given in ref.~\cite{YM}.
We introduce the Wilsonian effective action $\se[\phi,\g;\L]$,
define its relevant and irrelevant parts, discuss the r\^ole of
the boundary conditions and the local symmetry.
In section $3$ we study the operator which gives the violation to
the ST identities and show that the problem of implementing the
symmetry can be reduced to the solution of a finite number of equations,
the {\it fine tuning equations}.
Section $4$ contains the explicit solution of these fine tuning
equations and in
section $5$ we rephrase this solution in the usual algebraic language.
Section $6$ contains some conclusions.

\section{Renormalization group flow for $SU(2)$ Yang-Mills theory}
The fields and corresponding sources
for $SU(2)$ Yang-Mills theory are
$$
\phi=\{\,A^a_\mu,\,c^a,\,\bc^a \,\}
\, ,
\;\;\;\;\;\;\;\;
j=\{\, j^a_\mu,\, \bchi^a -\frac 1 g \partial_\mu u_\mu^a,\, - \chi^a \,\}
\, ,
\;\;\;\;\;\;\;\;
\g=\{u_\mu^a,\, v^a\}
\, ,
$$
where $u_\mu^a$ and $v^a$ are the sources associated to the BRS
variation of $A_\mu^a$ and $c^a$ respectively.
The generating functional, in the Feynman gauge, is
\beq\nome{W}
Z[j,\g]=e^{iW[j,\g]}=\int {\cal D}\phi \, \exp{i\{-\half (\phi p^2
\phi)_{0\L_0}+(j\phi)_{0\L_0}+\si[\phi,\g;\L_0]\}}
\,,
\eeq
where the path integral is regularized by assuming a UV cutoff $\L_0$.
In general we introduce the cutoff scalar products between fields and
sources
$$
\half (\phi \, p^2 \, \phi)_{\L\L_0}\equiv
\int_p\, p^2 \, K^{-1}_{\L\L_0}(p)\;
\left\{\half A^a_\mu(-p)\,A^a_\mu(p)
- \bc^a(-p)\, c^a(p)\right\}\,,
\;\;\;\;\;\;
\int_p \equiv \int \frac{d^4p}{(2\pi)^4}
$$
$$
(j\phi)_{\L\L_0}\equiv
\int_p \, K^{-1}_{\L\L_0}(p) \;
\left\{j^a_\mu(-p)\, A^a_\mu(p) +
[\bchi^a(-p) - \frac i g p_\mu u^a_\mu(-p)] \, c^a(p) +
\bc^a(-p)\, \chi^a(p) \right\}
\,,
$$
where $K_{\L\L_0}(p)$ is the cutoff function which is one
for $\L^2 < p^2 < \L_0^2$ and rapidly vanishes outside.

The functional $\si$ is the UV action involving monomials
in the fields, BRS sources and their derivatives which have dimension
not larger than four and are Lorentz and $SU(2)$ scalars. There are nine
of these independent monomials
\beeqn
&&
\si[\phi,\g;\L_0]
=\int d^4x \biggl\{
\frac 1 2 A_\mu \cdot
\left[
g_{\mu\nu}(
\s_1^B
+\s_2^B \partial^2)
+\s_3^B \partial_\mu \partial_\nu
\right] A_\nu
+\s_4^B\; w_\mu \cdot \partial_\mu c
\\&&\;\;\;\;\;\;
+\s_5^B\;(\partial_\nu A_\mu) \cdot A_\nu \wedge A_\mu
+\s_6^B \,(A_\mu\wedge A_\nu)\cdot (A_\mu\wedge A_\nu)\,
+\s_7^B \,(A_\mu\cdot A_\nu)\,(A_\mu\cdot A_\nu)\,
\\&&\;\;\;\;\;\;
+\s_8^B \;w_\mu\cdot c\wedge A_\mu
+\s_9^B \, v\cdot c \wedge c
\biggr\}\,
\,,
\eeeqn
where $w_\mu^a=u_\mu^a+g\partial_\mu \bc^a$ and we have introduced
the usual scalar and external $SU(2)$ product.
The nine couplings $\s_i^B$ depend on $\L_0$ and have non-negative
dimension (relevant parameters).
In order to obtain the physical theory one has to show that the values
of the $\s_i^B$ can be fixed in such a way that:
\newline
(1) the \UV limit can be taken by fixing the physical parameters such
as the masses, coupling $g$ and wave function normalization
constant at a subtraction point $\mu$.
Perturbative renormalizability ensures that this can be done
\cite{P,B,YM}.
\newline
(2) in the \UV limit the Slavnov-Taylor identities must be satisfied.
This is the crucial point to be discussed in this paper.

According to Wilson one integrates over the fields with frequencies
$\L^2<p^2<\L_0^2$ and obtains (see appendix A)
\beq\nome{Z}
e^{iW[j,\g]}=N[j,\g;\L] \;\int {\cal D}\phi \,
\exp i\{
-\half (\phi\, p^2 \phi)_{0\L}+(j\phi)_{0\L}+\se[\phi,\g;\L]
\; \}
\,,
\eeq
where $N[j,\g;\L]$ contributes to the quadratic part of $W[j,\g]$
\beq\nome{N}
\log N[j,\g;\L] = -i\int_p \frac 1{p^2}
\frac{K_{\L\L_0}(p)}{K_{0\L_0}(p) K_{0\L}(p)}
\left\{\half j^a_\mu(-p)\, j^a_\mu(p) -
[\bchi^a(-p) - \frac i g p_\mu u^a_\mu(-p)] \, \chi^a(p)
\right\}
\,.
\eeq
The functional $\se$ is the Wilsonian effective action and is obtained
by integrating the fields $\phi'$ with the higher modes  ($p^2 >\L^2$)
\beq\nome{esse}
\exp i\{\half (\phi \,p^2 \phi)_{\L\L_0} + \se[\phi,\g;\L]\}
=
\int {\cal D}\phi' \,
\exp i\{-\half (\phi' \,p^2 \phi')_{\L\L_0}+
(j' \phi' )_{\L\L_0}+\si[\phi',\g;\L_0]\} \,,
\eeq
where the source is
\beq\nome{j'}
j'(p)=p^2 \,\{\,A^a_\mu(p), \, -\bc^a(p), \, c^a(p)\,\}
\,.
\eeq
By comparing with the physical generating functional $W[j,\g]$
in \re{W} we see that
\beq\nome{WL}
W[j',\g;\L] \equiv
\half (\phi \, p^2 \phi)_{\L\L_0} + \se[\phi,\g;\L]\,,
\eeq
as a functional of $j'$, generates the connected Green functions in
which the internal propagators have frequencies in the range
$\L^2<p^2<\L_0^2$.
The cutoff function in the external propagators is cancelled
by the inverse cutoff so that the source $j'$ can be taken for all
values of $p$.
At \IR and \UV, this functional becomes
the physical generating functional
\beq\nome{phys}
W[j',\g]=W[j',\g;\L=0]=
\half (\phi \, p^2 \phi)_{0\L_0} + \se[\phi,\g;\L=0]
\,.
\eeq
Given the relation in \re{j'} between $j'$ and $\phi$ we have that the
vertices of $\se[\phi,\g;\L=0]$ are the amputated connected Green
functions.

Taking into account that the variable $\L$ enters as a cutoff in
the internal propagators of connected Green functions, one derives
the exact RG equation \cite{P}
\beq\nome{rg}
\dL \frac{\se}{\hbar} = (2\pi)^8\hbar
\int_p \frac {1}{p^2}\, \dL K_{0\L}(p) \;
e^{-i\set/\hbar}\;
\biggl\{
\half \frac{\de}{\de A_\mu^a(-p)}\frac{\de}{\de A_\mu^a(p)}
+
\frac{\de}{\de c^a(-p)}\frac{\de}{\de\bc^a(p)}
\biggr\}
\;e^{i\set/\hbar}\,.
\eeq
This equation is non-perturbative but it can be solved perturbatively
once the boundary conditions are given at some $\L$.
We have included $\hbar$ in \re{rg} to show explicitly how
the loop expansion is generated. In the r.h.s. there are two
contributions. One is quadratic in the action and has a coefficient
with the same power of $\hbar$ as the l.h.s, thus involving Green
functions with lower or equal loop order.
The second, linear in the action, has a coefficient with
an additional factor $\hbar$ and
therefore involves Green functions with a lower loop.
Thus, starting at zero loop with $\se^{(\ell=0)}$ given by the
value of Green functions at the boundary conditions, eq.~\re{rg} allows
one to obtain $\se^{(\ell)}$ at all loops.
The fact that one of the contributions in the r.h.s. in \re{rg} is at
the same loop order as the l.h.s. makes less transparent and less
efficient the generation of the loop expansion.
Actually, as shown in ref.~\cite{BDM,Wett}, the loop expansion is more
easily generated if one integrates the RG flow not for the Wilsonian
action $\se$ but for a cutoff effective action $\G[\phi,\g;\L]$
given by the usual Legendre transform of the cutoff functional
$W[j,\g;\L]$ introduced in \re{WL}
$$
\G[\phi,\g;\L]=W[j,\g;\L]-W[0,0;\L]-\int_p \, \left\{
j_\mu^a(-p)A_\mu^a(p)+\bchi^a(-p)c^a(p)+\bc^a(-p)\chi^a(p)\right\}
\,.
$$
This functional of the ``classical'' fields $\phi$ generates cutoff
vertex functions in which the internal propagators have frequencies in
the range $\L^2<p^2<\L_0$. From \re{phys} we have indeed that at $\L=0$
\beq\nome{G}
\G[\phi,\g]=\G[\phi,\g;\L=0]
\eeq
is the physical effective action.
One can convert \re{rg} into the RG equation for $\G[\phi,\g;\L]$
which has the following form
\beq\nome{rgg}
\dL \G[\phi,\g;\L]= \hbar I_{\G}[\phi,\g;\L]\,,
\eeq
where the functional $I_{\G}[\phi,\g;\L]$ is given in ref.~\cite{YM} and
depends non-linearly on the vertices of $\G[\phi,\g;\L]$.
Then in the r.h.s. of \re{rgg} there are vertices at a loop order lower
than the l.h.s., so that by solving iteratively this equation one
automatically generates the loop expansion.

We now discuss the crucial point of the boundary conditions which
provide the starting point for the loop expansion.

\subsection{Boundary conditions: physical parameters and symmetry}
There is an obvious value where the boundary conditions should
be given which is the UV value $\L=\L_0$.
Here one finds from \re{esse}
$$
\se[\phi,\g;\L=\L_0] =\si \,,
$$
thus $\se$ becomes local and depends only on the nine relevant couplings
$\s_i^B$ with non-negative dimension.
Since $\L_0$ is the only surviving mass parameter as \UV one expects
that, for \UV, all vertices with negative dimension vanish.
For this reason vertices with negative dimension
are called ``irrelevant'' and their contributions to the Wilsonian
action will be denoted by $\sei[\phi,\g;\L]$.

The remaining part
$\ser[\phi,\g;\s_i(\L)]$
is a local functional which contains nine relevant parameters
$\s_i(\L)$ with non-negative dimension.
The form of $\ser[\phi,\g;\s_i(\L)]$ is the
same as $\si$ in which the UV couplings $\s_i^B$ are replaced by
$\s_i(\L)$.
Also for these parameters we need to fix the boundary conditions at
some $\L=\L'$.
The precise definition of $\s_i(\L)$ in terms of $\se$ depends
crucially on whether one assumes boundary conditions at
$\L=\L' \ne 0$ (\eg $\L'=\L_0$) or at the physical point $\L'=0$.
In the first case the definition of $\s_i(\L)$ is simple since one
does not need to impose a subtraction point and can
consider the Taylor expansion of the Green functions around vanishing
momenta. Therefore one defines
$$
\ser[\phi,\g;\s_i(\L)]=
T^{(0)}_4 \, \se [\phi,\g;\L] \,,
\;\;\;\;\;
\L \ne 0 \,,
$$
where $T^{(0)}_4$ is the Taylor expansion around vanishing momenta
truncated to the terms with coefficients with non-negative dimension
(see for instance \cite{B}).

Taking advantage of the fact that
$\se[\phi,\g;\L=0]$ and $\G[\phi,\g;\L=0]$
are physical functionals (see \re{phys} and \re{G}),
we have suggested in ref.~\cite{YM} to fix the relevant parameters
$\s_i(\L)$ at the physical point $\L=0$ so that $\s_i(\L=0)$ are
physical couplings, such as for instance the mass, the wave function
normalizations and the coupling constant $g$ at a subtraction point
$\mu$.
At $\L=0$ the definition of the functional
$\ser[\phi,\g;\s_i(\L=0)]$
requires the introduction of a finite subtraction point $\mu$.
We then define
$$
\ser[\phi,\g;\s_i(\L)]=
T^{(\mu)}_4 \, \se [\phi,\g;\L] \,,
$$
where the operator $T^{(\mu)}_4$ is given in appendix B for
a scalar theory. The extension to the YM theory is simple in
principle and $T^{(\mu)}_4 \G[\phi,\g;\L]$
is defined completely in appendix B.

For the boundary conditions for the RG equation \re{rgg} of the
cutoff effective action $\G[\phi,\g;\L]$ we follow the same procedure:

\noindent
(1) we assume that the irrelevant parts of $\G[\phi,\g;\L=\L_0]$ vanish
$$
\Gi[\phi,\g;\L=\L_0]=
\left[1-T^{(\mu)}_4\right] \, \G[\phi,\g;\L=\L_0]=0 \,,
$$
(2) we fix the remaining nine relevant parameters at the physical
point $\L=0$.
It is useful to rearrange the various monomials and use the
following parametrization for the relevant part of the physical effective
action $\G[\phi,\g]=\G[\phi,\g;\L=0]$
\beq\nome{Grel}
\eqalign{
&
\Gr[\phi,\g]
=\int d^4x \biggl\{
-z_1\frac 1 4 F_{\mu\nu}\cdot F_{\mu\nu}
+z_2 \;
\left[
\frac 1 g w_\mu \cdot D_\mu c - \frac 1 2 \, v\cdot c \wedge c
\right]
\cr&
+g(z_3-1)\;(\partial_\nu A_\mu) \cdot A_\nu \wedge A_\mu
-
\frac 1 2 (\partial_\mu A_\mu) \cdot (\partial_\nu A_\nu)
\biggr\}\,
\;\;
+\Grt[\phi,\g;\r_i]\,,
}
\eeq
where
$F_{\mu\nu}=\partial_\mu A_\nu- \partial_\nu A_\mu +
g A_\mu \wedge A_\nu$,
$\;D_\mu c =\partial_\mu c +gA_\mu \wedge c$ and
\beq\nome{tG}
\eqalign{
&
\Grt[\phi,\g;\r_i] \equiv
\int d^4x \biggl\{
\r_1 \, \frac 1 2 A_\mu \cdot A_\mu
+\r_2 \,\frac 1 2 (\partial_\mu A_\mu) \cdot (\partial_\nu A_\nu)
+\r_3\, w_\mu \cdot c\wedge A_\mu
\cr&
+\frac 1 2 \r_4 \, v\cdot c \wedge c
+\frac{g^2}{4}\r_5\, (A_\mu\wedge A_\nu)\cdot(A_\mu\wedge A_\nu)
+\frac{g^2}{4} \r_6 \,(A_\mu\cdot A_\nu)\,(A_\mu\cdot A_\nu)\,
\biggr\}\,
\,.
}
\eeq
In \re{Grel} we have singled out two BRS invariant monomials, the
gauge fixing contribution and an additional three vector coupling monomial.

We now discuss the values $z_i$ and $\r_i$ that one has to fix in
order to have the physical couplings and the local $SU(2)$ gauge symmetry.
Some of these couplings can be fixed to give the wave function
normalizations and the three vector coupling $g$ at a subtraction
point $\mu$.
This can be implemented by assuming $z_1=z_2=z_3=1$ so that
the first part of $\Gr$ in \re{Grel} is just the BRS
action in the Feynman gauge $S_{\mbox{\footnotesize{BRS}}}[\phi,\g]$
and we have
\beq\nome{zi}
\Gr[\phi,\g]=S_{\mbox{\footnotesize{BRS}}}+\hbar \Grt[\phi,\g;\r_i]
\,.
\eeq
Notice that the choice of boundary conditions at $\L=0$ for the
relevant couplings provides a conspicuous advantage with
respect to the case in which one takes $\L=\L' \ne 0$.
We have in \re{zi} that the relevant part of the effective action
already contains the BRS invariant classical action which is the
starting point of perturbation theory. We have also explicitly written the
$\hbar$ dependence which shows that the other six couplings $\r_i$
vanish at zero loop.

In ref.~\cite{YM} we proposed to fix the $\r_i$'s from the ST identities
for the physical effective action $\G[\phi,\g]$
given by
\beq\nome{ST}
\D_\G[\phi,\g] \equiv {\cal S}_{\G'}\, \G'[\phi,\g]
=0
\,,
\eeq
where
$$
\G'[\phi,\g]=\G[\phi,\g]+
\frac 1 2 \int d^4x (\partial_\mu A_\mu)^2
$$
and ${\cal S}_{\G'}$ is the usual Slavnov operator \cite{BRS}
\beq\nome{SG}
{\cal S}_{\G'}=\int_p \biggl\{
\frac{\de\G'}{\de u^a_\mu(-p)}\frac{\de}{\de A^a_\mu(p)} +
\frac{\de\G'}{\de A^a_\mu(p)}\frac{\de}{\de u^a_\mu(-p)} +
\frac{\de\G'}{\de v^a(-p)}\frac{\de}{\de c^a(p)}+
\frac{\de\G'}{\de c^a(p)}\frac{\de}{\de v^a(-p)}
\biggr\}\,.
\eeq
The six couplings $\r_i$ are involved in all vertices of the
functional $\D_\G[\phi,\g]$.
In ref.~\cite{YM} we obtained the various $\r_i$ from the simplest ST
identities and found that these couplings are given by some
irrelevant part of vertices evaluated at the subtraction point $\mu$.
We showed that this procedure can be implemented perturbatively. This
is due to the fact that, as indicated in \re{zi}, the couplings $\r_i$
vanish at zero loop.
At higher loops some of the couplings $\r_i$ are different from zero.
The mass of the vector which is given by $\r_1$, as expected,
remains zero at all loops.
In that paper we were not able to show that this procedure
ensures that the ST identities are actually satisfied for all
vertices.
In the next section we will show that this is the case.

\section{Gauge symmetry and fine tuning equation at $\L=0$}
The proof that gauge symmetry can be implemented perturbatively
in the RG formulation has been given by Becchi in \cite{B}
in the case in which the boundary conditions for the nine parameters
$\s_i(\L)$ are given at some non-physical point $\L=\L'\ne 0$.
These parameters are defined by applying
the operator $T_4^{(0)}$ to the Wilsonian effective action.
The perturbative analysis of the ST identities is simplified by the
fact that $T_4^{(0)}$ commutes with the ST functional derivative
at zero loop.

Here we follow the same procedure but in the case in which one
fixes the boundary conditions at the physical point $\L=0$.
The price for working directly at $\L=0$ is that the relevant couplings
are defined by the operator $T_4^{(\mu)}$ which does not commute
with the zero loop ST operator.
As a result the analysis of the ST identities seems more complex.
We are able to show that actually the analysis is easy and one
can write a simple fine tuning equation which allows one to
determine the six relevant couplings $\r_i$.

Following Becchi we consider the generalized BRS transformation
\beeqn
&& \de A_\mu^a(p)=-\frac i g p_\mu \eta c^a(p) +K_{0\L}(p)\eta\,
\frac{\de \se}{\de
u_\mu^a(-p)}\,, \\
&& \de c^a(p)=K_{0\L}(p)\eta \,\frac{\de \se}{\de v^a(-p)}\,, \\
&& \de \bc^a(p)=\frac i g \eta p_\mu A_\mu^a(p)\,,
\eeeqn
with $\eta$ a Grassmann parameter.
Applying this change of variables to the functional integral \re{Z},
one deduces the following identity
$$
{\cal S}\,Z[j,\g]=
N[j,\g;\L]\,
\int {\cal D}\phi \; e^{i\{-\frac 1 2
(\phi p^2 \phi)_{0\L}+(j\phi)_{0\L}+\set[\phi,\g;\L]\}}
\; \D[\phi,\g;\L]\,,
$$
where
$$
{\cal S}\equiv \int_p \left( j^a_\mu(p)\frac{\de}{\de u^a_\mu(p)}
-\bchi^a(p) \frac{\de}{\de v^a(p)}
-\frac i g p_\mu \chi^a(p)\frac{\de}{\de j^a_\mu(p)} \right)
$$
is the usual ST operator obtained from the variation
of the source term $(j\phi)_{0\L}$. Notice that the cutoff
functions cancel.
The functional $\D$ arises from the variation of
the other terms in the exponent and from the Jacobian.
It can be divided into two parts, a linear and a quadratic one
in the derivatives
\beq\nome{D}
\D=\D_1+\D_2\,,
\eeq
with
$$
\D_1=i \int_p \biggl\{ p^2 A^a_\mu(p)\frac{\de}{\de u^a_\mu(p)}
+\frac i g p_\mu c^a(p)\frac{\de}{\de A^a_\mu(p)}
+\frac i g p_\mu w^a_\mu(p)
\frac{\de}{\de v^a(p)}
-\frac i g p_\mu A^a_\mu(p)\frac{\de}{\de \bc^a(p)} \biggr\} \se
$$
and
$$
\D_2=\int_p K_{0\L}(p) e^{-i\set}\,
\biggl\{
\frac{\de}{\de A^a_\mu(p)}\frac{\de}{\de u^a_\mu(-p)} +
\frac{\de}{\de c^a(p)}\frac{\de}{\de v^a(-p)}
\biggr\}\,
e^{i\set}\,.
$$
Since $K_{0\L}(p)$ vanishes for $\L=0$ we have that at the physical point
$\D[\phi,\g;\L=0]=\D_1[\phi,\g;\L=0]$.

We have to prove that one can select the six relevant couplings
$\r_i$ in such a way that
\beq\nome{D=0}
\D[\phi,\g;\L]=0
\,,
\eeq
so that the ST identities
$$
{\cal S}\,Z[j,\g]=0
$$
are satisfied.
Actually we will show that the functional equation \re{D=0} can be reduced
to a finite number of relations, the {\it fine tuning equations}.

The RG flow for the functional $\D$ is given by the following linear
evolution equation
\beq\nome{Dflow}
\dL \D[\phi,\g;\L]= M \cdot\D [\phi,\g;\L] \equiv M[\D;\L]\,,
\eeq
where the linear operator $M$ depends on $\se$
$$
M=-(2\pi)^8\int_p \frac{1}{p^2} \dL K_{0\L}(p)
\biggl\{\left(
\frac{\de\se}{\de A^a_\mu(p)} -\frac {i\hbar} {2} \frac{\de}{\de A^a_\mu(p)}
\right)\frac{\de}{\de A^a_\mu(-p)}
$$
$$
+\left( \frac{\de\se}{\de c^a(p)}-\frac{i\hbar}{2}\frac{\de}{\de c^a(p)}
\right)\frac{\de}{\de \bc^a(-p)}\;\; -\;\; c\leftrightarrow \bc\;\;
\biggr\}\,.
$$
The RG equation \re{Dflow} requires boundary conditions.
As before we discuss the boundary conditions for the relevant
part of $\D[\phi,\g;\L]$ at $\L=0$ and for the irrelevant part
at the UV point $\L=\L_0$.

The definition of the relevant part of $\D[\phi,\g;\L]$ at $\L=0$
requires a subtraction point. This functional has dimension $1$ and we have
\beq\nome{Drel}
\Dr[\phi,\g;\de_i]=T^{(\mu)}_5\,\D[\phi,\g;\L=0]
\,,
\eeq
where $T_5^{(\mu)}$ is again defined as in appendix B.
There are 11 relevant parameters $\de_i$ (see appendix C) which are
the coefficients of the 11 monomials in the fields, sources and
momenta of dimension not greater than five and with ghost number one.
Recall that at $\L=0$ the expression of $\D$ simplifies since
the non-linear part $\D_2$ vanishes due to the cutoff function.

We can easily prove that the irrelevant part of the functional $\D$
vanishes at $\L=\L_0$ provided that the UV cutoff is sent to infinity
\beq\nome{Dirr}
\Di[\phi,\g;\L=\L_0] \to 0\,,
\;\;\;\;\;\;
\mbox{for }
\;\;\;\;\;\;
\L_0 \to \infty
\,.
\eeq
First observe that $\D_1[\L=\L_0]$ contains only relevant parts.
This is due to the fact that at $\L=\L_0$ we have $\se[\L=\L_0]=\si$
which is local.
For $\D_2[\L=\L_0]$ we have that the functional in the
integrand is local apart from the $p$-dependence of the cutoff function
$K_{0\L_0}(p)$ which gives an irrelevant
contribution to $\D_2[\L=\L_0]$.
However, for \UV we have $K_{0\infty}(p)=1$ so that
$\D_2[\L=\L_0\to \infty]$ becomes local.

Using the boundary condition \re{Drel} and \re{Dirr} we can integrate
the RG equation for $\D$ and obtain
\beq\nome{inteq}
\D[\phi,\g;\L]=
\Dr[\phi,\g;\de_i]+
\int_0^\L \frac{d\l}{\l}
M_{\mbox{\footnotesize{rel}}}[\D;\l]-
\int_\L^\infty \frac{d\l}{\l}
M_{\mbox{\footnotesize{irr}}}[\D;\l]\,.
\eeq
We now prove the following two perturbative theorems
for the functional $\D^{(\ell)}[\phi,\g;\L]$ at loop $\ell$ and for
the corresponding vertices $\D_n^{(\ell)}(...;\L)$ where $n$ denotes the
number of fields or sources and the dots denote momenta, internal and
Lorentz indices.

\noindent
Theorem 1:
if $\D^{(\ell')}[\phi,\g;\L]=0$ for all loop $\ell'<\ell$, and if
$\D^{(\ell)}_{n'}(...;\L)=0$ for $n'<n$,
then
$$
\D^{(\ell)}_{n}(...;\L)= \D^{(\ell)}_{\mbox{\footnotesize{rel}},n}(...;\L=0)
\,.
$$
Proof: from the hypothesis the functional $M[\D;\L]$ at loop $\ell$ is
given by
\beeq\nome{Mell}
&& M^{(\ell)}[\D;\L]=-(2\pi)^8\int_p \frac{1}{p^2} \dL K_{0\L}(p)
\\ \nonumber && \;\;\;\;\;\;
\times \left(
\frac{\de\se^{(0)}}{\de A^a_\mu(p)}\frac{\de}{\de A^a_\mu(-p)}
+\frac{\de\se^{(0)}}{\de c^a(p)}\frac{\de}{\de \bc^a(-p)}
\;-\; c\leftrightarrow \bc\;\right)\D^{(\ell)}[\phi,\g;\L]\,,
\eeeq
where $\se^{(0)}$ is the tree Wilsonian action in \re{WL}.
Since the starting point of the perturbative expansion (see \re{zi})
is $\G^{(0)}[\phi,\g;\L]=S_{\mbox{\footnotesize{BRS}}}[\phi,\g]$,
the vertices of $\se^{(0)}[\phi,\g;\L]$ are the amputated connected Green
functions at tree level. Moreover $\se^{(0)}[\phi,\g;\L]$ does not contain
contributions quadratic in the fields, they are all included in
$\half (\phi p^2\phi)$, see \re{WL}.
Consider now the RG equations \re{inteq} for a vertex
$\D_n^{(\ell)}(...;\L)$.
{}From \re{Mell} we have that $M^{(\ell)}[\D;\L]$
involves only vertices $\D_{n'}^{(\ell)}(...;\l)$ with $n'<n$.
They vanish according to the hypothesis thus only the first term in
r.h.s. of \re{inteq} is left and the theorem is proved.
An obvious consequence of this is the second theorem.

\par\noindent
Theorem 2:
if $\D^{(\ell')}[\phi,\g;\L]=0$ for all loop $\ell'<\ell$
and if at loop $\ell$ the relevant part of this functional
vanishes at $\L=0$, namely if
\beq\nome{ft}
\Dr[\phi,\g;\de^{(\ell)}_i]=0\,,
\eeq
then $\D^{(\ell)}[\phi,\g;\L]=0$.

The fine tuning condition \re{ft} will be used to fix the
relevant couplings $\r_i$ of the effective action $\G[\phi,\g]$.
However the functional $\D[\phi,\g;\L]$ defined in \re{D}
involves vertices of the Wilsonian action $\se[\phi,\g;\L]$.
It is convenient to express the fine tuning condition
at $\L=0$ in terms of a functional which involves directly the
vertices of $\G[\phi,\g]$.
This can be simply done by taking the Legendre transform
of $\D[\phi,\g;\L]$ at $\L=0$.
In this way the fine tuning equation \re{ft} is equivalent to
\beq\nome{ft'}
\DGr[\phi,\g;\de_i^{(\ell)}]=T_5^{(\mu)}
\D_\G^{(\ell)}[\phi,\g]=0\,,
\;\;\;\;\;\;\;\;\;\;
\D_\G[\phi,\g]\equiv {\cal S}_{\G'} \G'[\phi,\g]\,,
\eeq
where $\DGr$ has the same decomposition as $\Dr$ given in appendix C.
Then both theorems 1 and 2 can be phrased for the functional $\D_\G[\phi,\g]$.

{}From now on we consider $\L=0$ and we show perturbatively the equivalence of
\re{ft} and \re{ft'}.
By using the inductive assumption $\D^{(\ell')}=0$ for $\ell'<\ell$,
one has the following results at loop $\ell$:

\noindent
(1) For the two field vertex one has
$\D^{(\ell)(Ac)}_\G=\D^{(\ell)(Ac)}$, then the fine tuning condition
$\Dr^{(\ell)(Ac)}=0$ is equivalent to
$\DGr^{(\ell)(Ac)}=0$. Imposing this condition the full vertex
$\D^{(\ell)(Ac)}_\G$ vanishes.

\noindent
(2) For vertices with three fields one has
$\D^{(\ell)(AAc)}_\G=\D^{(\ell)(AAc)}$ and
$\D^{(\ell)(wcc)}_\G=\D^{(\ell)(wcc)}$, since the one particle
reducible terms vanish due to (1).
Thus the fine tuning conditions
$\Dr^{(\ell)(AAc)}=0$ and $\Dr^{(\ell)(wcc)}=0$ are
equivalent to $\DGr^{(\ell)(AAc)}=0$ and
$\DGr^{(\ell)(wcc)}=0$, respectively. With these
conditions these two vertices vanish.

\noindent
(3) Similarly for the vertices with four fields one has
$\D^{(\ell)(3Ac)}_\G=\D^{(\ell)(3Ac)}$ and
$\D^{(\ell)(wAcc)}_\G=\D^{(\ell)(wAcc)}$ and the remaining
fine tuning conditions
become $\DGr^{(\ell)(3Ac)}=0$ and
$\DGr^{(\ell)(wAcc)}=0$ and the two functions
$\D^{(\ell)(3Ac)}_\G$ and $\D^{(\ell)(wAcc)}_\G$ vanish.

\noindent
(4) By increasing the number of fields one has that
the two functionals $\D^{(\ell)}$ and $\D^{(\ell)}_\G$ are equal.

In the next section we prove by induction on the number of loops that
it is possible to fix the six couplings $\r_i^{(\ell)}$ in such a way
that $\de^{(\ell)}_i=0$, so that the fine tuning equation is solved and
from theorems 1 and 2 the ST identities hold perturbatively.

\section{Perturbative solution of the fine tuning equation}
The starting point of the proof by induction over the number of loops
is $\D_\G^{(\ell=0)}[\phi,\g]=0$ which is valid since
$\G^{(\ell=0)}=S_{\mbox{\footnotesize{BRS}}}$.
We suppose that the fine tuning equation is solved at $\ell'<\ell$
so that $\D^{(\ell')}_\G[\phi,\g]=0$. We want to show that it is
possible to fix the six couplings $\r^{(\ell)}_i$ in such a way
that $\D^{(\ell)}_\G[\phi,\g]=0$.
Notice that there are $6$ parameters $\r_i^{(\ell)}$ and $11$
equations $\de_i^{(\ell)} =0$.
Therefore the solvability of these equations requires that there
must be $5$ relations connecting the various $\de_i^{(\ell)}$.
They are provided by the so-called consistency conditions, a set of
equations which must be identically satisfied by the functional
$\D_\G$ due to its definition \re{ST}. From the
anticommuting character of ${\cal S}_{\G'}$ we have
${\cal S}_{\G'}\,\D_\G={\cal S}^2_{\G'}\,\G'=0$,
where ${\cal S}_{\G'}$ is defined in \re{SG}.
By using again the inductive hypothesis $\D_\G^{(\ell')}=0$ for $\ell'<\ell$,
the consistency condition simplifies and at loop $\ell$ one finds
\beq\nome{cc}
{\cal S}_{\G_0'}\,\D^{(\ell)}_\G=0\,,
\eeq
where ${\cal S}_{\G_0'}$ is defined in \re{SG} with $\G$ given by
$S_{\mbox{\footnotesize{BRS}}}$.
In order to obtain from this equation the desired relations connecting
the various $\de^{(\ell)}_i$, we have to extract
its relevant part by applying the operator $T^{(\mu)}_6$.
Notice that this operator mixes relevant and irrelevant
parts of $\D_\G$. This in general makes some of the consistency
conditions complicated due to the presence of irrelevant contributions.
We analyse in detail the fine tuning equations and consistency
conditions for the various vertices of $\D_\G$.
We start form the vertices with two and three fields and sources.
We then analyse the remaining vertices.
In the following all vertices will be considered at loop $\ell$.

\noindent
(1) The vertices with two and three fields are obtained from the
definition of $\D_\G$ in \re{ST}
\beq\nome{Ac}
\D^{(Ac)}_{\G,\mu}(p)={\G'}_{\mu\nu}^{(AA)}(p)\,\G_\nu^{(wc)}(p)
\,,
\eeq
\beq\nome{AAc}
\D^{(AAc)}_{\G,\mu\nu}(p,q,k)=\G_\r^{(wc)}(k)\,\G_{\mu\nu\r}^{(3A)}(p,q,k)-
{\G'}_{\mu\r}^{(AA)}(p) \, \G_{\r\nu}^{(wcA)}(p,k,q)+
{\G'}_{\nu\r}^{(AA)}(q) \, \G_{\r\mu}^{(wcA)}(q,k,p)
\,,
\eeq
\beq\nome{wcc}
\D^{(wcc)}_{\G,\mu}(p,q,k)=\G_\mu^{(wc)}(p)\,\G^{(vcc)}(p,q,k)+
\G_\nu^{(wc)}(q) \, \G_{\mu\nu}^{(wcA)}(p,k,q)+
\G_\nu^{(wc)}(k) \, \G_{\mu\nu}^{(wcA)}(p,q,k)
\,.
\eeq
The relevant parameters of these vertices are the five
$\de_i$ with $i=1,2,3,4,5$ (see appendix C) while the relevant
couplings of the $\G$ vertices are the four $\r_i$ with
$i=1,2,3,4$ (see \re{Grel} and appendix B).
Then we need one relation among these $\de_i$.
Indeed if we consider the consistency condition \re{cc} for the
vertex of the $A$-$c$-$c$ fields and take its relevant part, we obtain
\beq\nome{cc234}
i g\de_2=\de_3+\de_{4}\,.
\eeq
It is a nice fact that this relation involves only relevant parameters.
This is not a general fact since the operator $T_6^{(\mu)}$ applied to
${\cal S}_{\G}\,\D_\G$ typically extracts irrelevant vertices of $\D_\G$
evaluated at the subtraction point.

We are now able to satisfy the fine tuning equation for the vertices
in \re{Ac}-\re{wcc}. From \re{Ac} we obtain
$$
\de_1=-\frac i g \r_1\,,
\;\;\;\;\;\;\;\;\;\;\;
\de_2=-\frac i g \r_2\,,
$$
and we have $\de_1=\de_2=0$ by fixing
\beq\nome{r12}
\r_1=\r_2=0\,.
\eeq
{}From \re{AAc} we obtain
$$
\de_3=\r_3\,,
$$
and we have $\de_3=\de_{4}=0$ by fixing
\beq\nome{r3}
\r_3=0\,.
\eeq
Finally from \re{wcc} we obtain
$$
\de_5=-\frac i g \left\{
\r_4 + \left[\frac{2}{p^2} p_\mu k_\nu
\tilde \G_{\mu\nu}^{(wcA)}(p,q,k) \right]_{3SP}
\right\}\,,
$$
and we have $\de_5=0$ by fixing
\beq\nome{r4}
\r_4=-\left[\frac{2}{p^2} p_\mu k_\nu
\tilde \G_{\mu\nu}^{(wcA)}(p,q,k) \right]_{3SP}\,.
\eeq
{}From theorem 2 of the previous section we have that,
since the relevant parts of the vertices in \re{Ac}-\re{wcc}
vanish, the complete vertices vanish.
Moreover from theorem 1 of previous section the four field vertices
of $\D_\G$ could be only relevant.

\noindent
2) We come now to analyse the vertices of $\D_\G$ involving
four fields. Consider first the vertex
$\D^{(wAcc)abcd}_{\G,\mu\nu}(p,q,k,h)$,
which contains the relevant parameter $\de_{6}$ (see appendix C).
{}From the definition of $\D_\G[\phi,\g]$ in \re{ST} one easily sees
that this vertex involves only couplings $\r_i$ already fixed thus one
should be able to show that the fine tuning condition $\de_{6}=0$ is
satisfied automatically.
Indeed there is a consistency condition which, after having imposed
\re{r12}, \re{r3} and \re{r4}, reduces to $\de_{6}=0$.
This is due to theorem 1 which implies that irrelevant parts of the
vertices of $\D_\G$ with four fields vanish.

The last vertex of $\D_\G$ we have to consider is
$$
\D^{(3Ac)abcd}_{\G,\mu\nu\r}(p,q,k,h)=
\G_\s^{(wc)}(h)\,\G_{\mu\nu\r\s}^{(4A)abcd}(p,q,k,h)
$$
$$
+{\G'}_{\mu\s}^{(AA)}(p) \, \G_{\s\nu\r}^{(wcAA)adbc}(p,h,q,k)
+{\G'}_{\nu\s}^{(AA)}(q) \, \G_{\s\mu\r}^{(wcAA)bdac}(q,h,p,k)
$$
\beq\nome{AAAc}
+{\G'}_{\r\s}^{(AA)}(k) \, \G_{\s\mu\nu}^{(wcAA)cdab}(k,h,p,q)
+\eps^{eda}\eps^{ebc} \G_{\s\mu}^{(wcA)}(q+k,h,p) \,
\G_{\s\nu\r}^{(3A)}(p+h,q,k)
\eeq
$$
+\eps^{edb}\eps^{eac} \G_{\s\nu}^{(wcA)}(p+k,h,q) \,
\G_{\s\mu\r}^{(3A)}(q+h,p,k)
+\eps^{edc}\eps^{eba} \G_{\s\r}^{(wcA)}(q+p,h,k) \,
\G_{\s\nu\mu}^{(3A)}(k+h,q,p)\,.
$$
This vertex contains the five relevant parameters $\de_i$ with
$i=7,\dots, 11$ while the $\G$ vertices involved here contain only
the two relevant couplings not yet fixed, $\r_5$ and $\r_6$.
The consistency condition \re{cc} gives the three relations
\beq\nome{cclast}
\de_7=\de_9=\de_{11}\,,
\;\;\;\;\;\;\;\;\;
\de_8=\de_{10}\,.
\eeq
Notice that these relations involve only relevant parameters.
As stated before this is a consequence of the fact
that we have already set to zero the vertices of $\D_\G$ with two and
three fields and, according to theorem 1, the vertices with four
fields do not have any irrelevant parts.

{}From \re{cclast} we have that it is possible to have
$\de_i=0, \; i=7,\dots, 11$, simply setting to zero two independent
$\de_i$ (\eg $\de_7$ and $\de_8$) by fixing $\r_5$ and $\r_6$.
The easiest way of obtaining $\de_7$ and $\de_8$ is to contract
in two different ways the vertex in eq.~\re{AAAc} by external momenta
and summing over internal indices.
Since this vertex does not have any irrelevant part, these
contractions are proportional to two different linear combinations of
$\de_7$ and $\de_8$.
Using for simplicity the symmetric point we find
\beeqn
\frac{p_\mu q_\nu k_\r}{p^4}
\D^{(3Ac)aabb}_{\mu\nu\r}(p,q,k,h)|_{4SP} &=&
-\frac{15}{9}(2 \de_7 + \de_8)\\ &=&
\;-\frac{10}{3}ig \r_6 -\frac i g \left[
\frac{p_\mu q_\nu k_\r h_\s}{p^4} \, \tilde \G^{(4A)aabb}_{\mu\nu\r\s}(p,q,k,h)
\right]_{4SP}
\eeeqn
and
$$
\frac{p_\mu q_\nu h_\r}{p^4}
\D^{(3Ac)aabb}_{\mu\nu\r}(p,q,k,h)|_{4SP} =
-\frac 1 3 (2\de_7-7\de_8)
$$
$$=
ig(\frac{16}{3}\r_5-\frac 2 3 \r_6)
-\biggl[\frac{i}{g}\frac{p_\mu q_\nu h_\r h_\s}{p^4}
\tilde \G_{\mu\nu\r\s}^{(4A)aabb}(p,q,k,h)
$$
$$
+\frac{p_\mu q_\nu}{p^2} (h_\r +\frac 1 3 k_\r)
\G_{\r\mu\nu}^{(wcAA)aabb}(k,h,p,q)
+12 \;\frac{p_\mu q_\nu h_\r}{p^4}
\G_{\s\mu}^{(wcA)}(q+k,h,p) \,
\G_{\s\nu\r}^{(3A)}(p+h,q,k)
\biggr]_{4SP}\,.
$$
The fine tuning equations $\de_7=0$ and $\de_8=0$ give
\beq\nome{r6}
\r_6 =-\frac{3}{10g^2} \left[ \frac{p_\mu q_\nu k_\r h_\s}{p^4}
\, \tilde \G^{(4A)aabb}_{\mu\nu\r\s}(p,q,k,h)
\right]_{4SP}
\eeq
and\footnote{In ref.~\cite{YM} some contributions to \re{r5} were
missing.}
\beeq\nome{r5}
&&\r_5=\frac{1}{8} \;\r_6-\frac{3i}{16g}\biggl[\frac{i}{g}
\frac{p_\mu q_\nu h_\r h_\s}{p^4} \tilde \G_{\mu\nu\r\s}^{(4A)aabb}(p,q,k,h)
+\frac{p_\mu q_\nu}{p^2} (h_\r +\frac 1 3 k_\r)
\G_{\r\mu\nu}^{(wcAA)aabb}(k,h,p,q) \nonumber
\\ &&\;\;\;\;\;\;\;\;\;\;\;\;
+12 \; \frac{p_\mu q_\nu h_\r}{p^4}
\G_{\s\mu}^{(wcA)}(q+k,h,p) \,
\G_{\s\nu\r}^{(3A)}(p+h,q,k)
\biggr]_{4SP}\,.
\eeeq

This completes the proof that we can set to zero the $11$ parameters
$\de_i$ by fixing, at every loop, the $6$ couplings $\r_i$ according
to \re{r12}, \re{r3}, \re{r4}, \re{r5} and \re{r6}.

\section{Algebraic formulation}
We want to connect the perturbative solution we have discussed with the
algebraic formulation given in ref.~\cite{B}.
This is based on the fact that
if $\DGr[\phi,\g;\de_i]$ can be written in the following form
\beq\nome{*1}
\DGr[\phi,\g;\de_i]=
{\cal S}_{{\G'}^{(0)}}\,
\Gr[\phi,\g;{\bar \r}_i]
\,,
\eeq
then it is possible to find a perturbative solution of the fine tuning
equation by appropriately fixing the $\r_i$.
This property is discussed in \cite{B} and recalled in appendix D.
Eq.~\re{*1} is valid provided the $11$ parameters in $\DGr$
fulfil the following five relations
\beq\nome{*2}
ig\de_2=\de_3+\de_{4}\,,
\eeq
\beq\nome{*3}
\de_{6}=-ig\de_5\,,
\;\;\;\;\;\;\;\;\;
\de_7=\de_9=\de_{11}\,,
\;\;\;\;\;\;\;\;\;
\de_8=\de_{10}\,.
\eeq
In our case, in which we have a non-vanishing subtraction point
$\mu\ne 0$, the first relation \re{*2} holds (see \re{cc234}),
while the relations \re{*3} are not valid in general
since there are contributions from irrelevant parts.

Still also for $\mu\ne 0$ we can use the above theorem as follows.
First observe that the relation \re{*2} holds also for $\mu\ne 0$
and involves only vertices with two and three fields, \ie
$\D_\G^{(Ac)}$, $\D_\G^{(AAc)}$ and  $\D_\G^{(wcc)}$.
Therefore the form \re{*1} is valid for these vertices and
from appendix D we deduce that we can solve the fine tuning equations
$\de_i=0$ for the 5 relevant parameters with $i=1,\dots, 5$.
This is obtained by fixing the four couplings
$\r_1$, $\r_2$, $\r_3$ and $\r_4$,
which enter in the vertices of $\G$ with two and three fields.
{}From the previous section the solution is
$\r_1=\r_2=\r_3=0$ and $\r_4$ given in \re{r4}.

After solving these five  fine tuning conditions,
we have from theorem 2 that the complete vertices
$\D_\G^{(Ac)}$, $\D_\G^{(AAc)}$ and  $\D_\G^{(wcc)}$ vanish
and from theorem 1 that the vertices with four fields are
only relevant.
Therefore the relations in \re{*3} are now valid
and all vertices of $\DGr[\phi,\g;\de_i]$
can be expressed as in \re{*1}.
{}From appendix D we can perturbatively solve the remaining
fine tuning equations and from previous section the solution is
$\r_5$ in \re{r5} and $\r_6$ in \re{r6}.

\section{Conclusions}
In this paper we have fixed the renormalization conditions for the
$SU(2)$ Yang-Mills theory,
\ie the nine couplings ($z_i$ and $\r_i$, see \re{Grel}),
which enter in the relevant part of the effective action $\G[\phi,\g]$
as boundary conditions at $\L=0$ for the exact RG flow.
Three of these couplings are fixed to give, at the subtraction point
$\mu \ne 0$,
the vector and ghost wave functions and the three vector
coupling ($z_1=z_2=z_3=1$).
In this way one fixes a contribution of the effective action to
be the BRS classical action (for instance in the Feynman gauge, see
\re{zi}).
The other six couplings $\r_i$ in \re{tG} are absent at tree level.
They are at our disposal in order to implement the gauge symmetry
for the physical effective action, \ie the ST identities
$\D_\G[\phi,\g]=0$
in \re{ST}.
The fact that the three couplings $z_i=1$ are not affected by loop
corrections is one of the advantages of working at the physical
point $\L=0$ rather than at $\L=\L' \ne 0$.
(We shall recall later the difficulties arising from the
non-vanishing subtraction point).
We have shown perturbatively the following two results.

The first result (see section 3) is that it is possible to satisfy the
ST identities $\D_\G[\phi,\g]=0$ if one is able to solve the fine tuning
equations $\DGr[\phi,\g;\de_i]=0$.
It is a consequence of the RG equation \re{Dflow} (see theorem 1 and 2)
and of the fact that the starting point of the loop expansion is
the classical BRS action which satisfies the ST identities
$\D^{(\ell=0)}_\G=0$.

The second result (see sections 4 and 5) is that it is possible
to solve (perturbatively) the eleven fine tuning equations $\de_i=0$
by fixing the six $\r_i$ couplings.
This is possible since, due to the consistency condition \re{cc},
there are only six independent $\de_i$.
Moreover we have constructed the solution and found that
the relevant part of the effective action is
$$
\Gr[\phi,\g;\r_i] =S_{\mbox{\footnotesize{BRS}}}+
\hbar
\int d^4x \biggl\{
\frac {\r_4}{2} \, v\cdot c \wedge c
+\frac{g^2\r_5 }{4}(A_\mu\wedge A_\nu)\cdot(A_\mu\wedge A_\nu)
+\frac{g^2\r_6 }{4} (A_\mu\cdot A_\nu)\,(A_\mu\cdot A_\nu)\,
\biggr\}.
$$
The only non-vanishing couplings $\r_4$, $\r_5$ and $\r_6$
are given in \re{r4}, \re{r5} and \re{r6}, respectively,
in terms of irrelevant vertices of $\G$ evaluated at the
subtraction point. This form allows one to deduce the perturbative
expansion since irrelevant vertices at loop $\ell$ involve
relevant couplings at lower loops $\ell'<\ell$.

We have followed the method used by Becchi \cite{B} in which the
boundary conditions are taken at the non-physical point $\L=\L' \ne 0$.
In that case one can define the relevant parameters by expanding
around vanishing momenta thus avoiding irrelevant
contributions in the consistency conditions.

Since we work at $\L=0$ we have needed non-vanishing subtraction points
and introduced the operator $T_k^{(\mu)}$ which defines the relevant
parts of the various functionals.
This makes the analysis of the ST identities more difficult since
the operator $T_k^{(\mu)}$ mixes relevant and irrelevant parts
when applied to the product of two functionals.
As discussed in section 5 the fine tuning equation
$\D_\G[\phi,\g;\de_i]=0$ can be perturbatively solved provided that the
local functional $\DGr$ can be parametrized as in \re{*1}.
This is valid if the relations \re{*2} and \re{*3} hold.
Actually the relations \re{*3} are not valid in general if $\mu \ne 0$ due
to the presence of the irrelevant contributions generated by the
$T_6^{(\mu)}$ operator.
However we have shown that one can use the parametrization in \re{*1}
by proceeding in two steps.
First one uses the fact that eq.~\re{*2} is valid also for $\mu \ne
0$. This allows one to solve the fine tuning equations for the vertices
of $\D_\G$ with two and three fields.
Once these equations are solved the remaining relations \re{*3}
hold since the irrelevant contributions vanish.
In this way one can use \re{*1} and solve the remaining fine tuning equations.

The method is general. As shown in ref.~\cite{chi} it can be applied for
instance to $SU(2)$ gauge theory with fermions.
The application to the case of chiral gauge theory without anomalies
should be also possible along the same lines.

\vspace{3mm}\noindent{\large\bf Acknowledgements}

We have benefited greatly from discussions with C. Becchi and M. Tonin.

\newpage

\noindent
{\large\bf Appendix A}
\vskip 0.3 true cm
\noindent
The generating functional \re{W} is equivalent to
\beq\nome{W12}
N[j,\g;\L] \;
\int {\cal D}\phi{\cal D}\phi_1 \,
\exp{i\{-\half (\phi p^2 \phi)_{0\L}
-\half (\phi_1 p^2 \phi_1)_{\L\L_0}
+(j\phi)_{0\L}+\si[\phi+\phi_1,\g;\L_0]\}}
\,,
\eeq
where
$$
K_{0\L_0}(p)\;= K_{0\L}(p)\;+\;K_{\L\L_0}(p)
$$
and the coefficient $N$ is given in eq.~\re{N}.
This can be easily seen by making in \re{W12} the change of
variables $\phi_1=\phi'-\phi$, which gives
\beq\nome{W'}\eqalign{
N[j,\g;\L] \;&
\int {\cal D}\phi'
\exp{i\{-\half (\phi' p^2 \phi')_{\L\L_0}+\si[\phi',\g;\L_0]\}}\;
\cr&
\times
\int {\cal D}\phi \,
\exp{i\{-\half (\phi p^2 \phi)_{0\L}
-\half (\phi p^2 \phi)_{\L\L_0}
+(j_1\phi)_{\L\L_0}
+(j\phi)_{0\L}\}}\,,
}
\eeq
where the source $j_1(p)$ is
$$
j_1(p)=p^2 \,(A'^a_\mu(p), \, -\bc'^a(p), \, c'^a(p)\,)\,.
$$
By performing the integral over the field $\phi$, which is
gaussian, one obtains \re{W}.
On the other hand by integrating over the field $\phi'$ in \re{W'}
and using \re{esse} one obtains \re{Z}.

\vskip 1 true cm

\noindent
{\large\bf Appendix B}
\vskip 0.3 true cm
\noindent
We explicitly give the form of the operator
$T_4^{(\mu)}$, which extracts the relevant part of a
functional of a multicomponent massless scalar field $\psi_i$.
Due to the masslessness of the field, the Taylor expansion around
vanishing momenta is
affected by infrared singularities when we consider the four
dimensional terms in the fields and momenta.
Thus one is forced to introduce a non-vanishing subtraction point.
For the two field components this point is assumed
at $p^2=\mu^2$,
while for the $N$ field components it is assumed at the symmetric point
$NSP$ defined by
$$
\bp_i\bp_j=\frac{\mu^2}{N-1}(N\de_{ij}-1)\,,\;\;\;\;\;\;\;\;N=3,4,...\,.
$$
In order to have a more compact notation, we also introduce the Fourier
transform $\tilde \psi_i$ of the field $\psi_i$. The form of
$T_4^{(\mu)}$ is then
\beeqn
T_4^{(\mu)}\,F[\psi] &\equiv& F[0]+\int d^4x \, \psi_i(x) \biggl\{
\bigl[\frac{\de F}{\de\psi_i(0)}\bigr]_{\psi=0} +
\frac 1 2 \psi_j(x) \bigl[
\frac{\de^2 F}{\de{\tilde \psi}_j(0)\de\psi_i(0)}\bigr]_{\psi=0} \\
&+& \frac i 2 \partial_\mu \psi_j(x) \bigl[ \partder{p_\mu}
\frac{\de^2 F}{\de{\tilde \psi}_j(p)\de\psi_i(0)}\bigr]_{p=0,\psi=0} +
\frac 1 6 \psi_j(x) \psi_k(x) \bigl[
\frac{\de^3 F}
{\de{\tilde \psi}_j(0)\de{\tilde \psi}_k(0)\de\psi_i(0)}\bigr]_{\psi=0}
\\ &-&\frac 1 2 \partial^2 \psi_j(x) \bigl[ \partder{p^2}
\frac{\de^2 F}{\de{\tilde \psi}_j(p)\de\psi_i(0)}\bigr]_{p^2=\mu^2,\psi=0}
\\ &+& \frac i 6 \psi_j(x) \partial_\mu \psi_k(x) \bigl[ \partder{p_{2\mu}}
\frac{\de^3 F}{\de{\tilde \psi}_j(p_1)\de{\tilde \psi}_k(p_2)
\de\psi_i(0)}\bigr]_{p_i=3SP,\psi=0}
\\ &+& \frac{1}{24} \psi_j(x) \psi_k(x) \psi_h(x) \bigl[
\frac{\de^4 F}
{\de{\tilde \psi}_j(p_1)\de{\tilde \psi}_k(p_2)\de{\tilde \psi}_h(p_3)
\de\psi_i(0)}\bigr]_{p_i=4SP,\psi=0} \,.
\eeeqn

In the following we explicitly give the relevant part of the
effective action $\G[\phi,\g]$ for the SU(2) Yang-Mills theory.
We first isolate the vertices of $\G[\phi,\g]$ which contain the
relevant couplings
\beq\nome{rel}\eqalign{
\G[\phi,\g] &= \frac 1 2 \int_p \G_{\mu\nu}^{(AA)}(p) A^a_\mu(-p)
A^a_\nu(p)+
\frac {1} {3!} \eps^{abc} \int_p \int_q \G_{\mu\nu\rho}^{(3A)}(p,q,r)
A^a_\mu(p) A^b_\nu(q) A^c_\rho(r) \cr
&+ \frac {1} {4!} \int_p \int_q \int_k
\G_{\mu\nu\r\s}^{(4A)abcd}(p,q,k,h)
A^a_\mu(p) A^b_\nu(q) A^c_\r(k) A^d_\s(h) \cr
&+ \int_p \G_{\mu}^{(wc)}(p) w^a_\mu(-p) c^a(p)
+ \eps^{abc} \int_p \int_q  \G_{\mu\nu}^{(wcA)}(p,q,r)
w^a_\mu(p) c^b(q) A_\nu^c(r) \cr
&+ \frac 1 2 \eps^{abc} \int_p \int_q
\G^{(vcc)}(p,q,r) v^a(p) c^b(q) c^c(r) +\ldots \,,
}
\eeq
where $r=-p-q$, $h=-p-q-k$ and the dots stand for
all the remaining terms which contain only irrelevant vertices,
since they are coefficients of monomials in the fields and sources
with dimension higher than four.

The six vertices in \re{rel} contain $9$ relevant couplings
which are defined as follows:
\beeqn
&&\G_{\mu\nu}^{(AA)}(p)=
g_{\mu\nu}[\s_{m_A}+p^2 (-1+\s_\alpha) + \Sigma_L(p)]
+ t_{\mu\nu}(p) [\s_A + \Sigma_T(p)]\,,
\\&&
\G_{\mu\nu\r}^{(3A)}(p,q,r)=
[(p-q)_\r g_{\mu\nu} +(q-r)_\mu g_{\nu\r} + (r-p)_\nu g_{\mu\r}]\;
[\s_{3A}+\S^{(3A)}(p,q,r)]
\\ && \;\;\;\;\;\;\;\;\;\;\;\;\;\;\;\;\;\;\;\;\;\;\;
+\tilde \G_{\mu\nu\r}^{(3A)}(p,q,r)\,,
\\&&
\G_{\mu\nu\r\s}^{(4A)abcd}(p,q,k,h)
=t^{abcd}_{1;\mu\nu\r\s}[\s_{4A}+\S_1^{(4A)}(p,q,k,h)]
+t^{abcd}_{2;\mu\nu\r\s}[\s'_{4A}+\S_2^{(4A)}(p,q,k,h)]
\\ && \;\;\;\;\;\;\;\;\;\;\;\;\;\;\;\;\;\;\;\;\;\;\;
+\tilde \G_{\mu\nu\r\s}^{(4A)abcd}(p,q,k,h)\,,
\\&&
\G_\mu^{(wc)}(p)=\frac{p_\mu}{g}[-i+\s_{wc}+\S^{(wc)}(p)]\,,
\\&&
\G_{\mu\nu}^{(wcA)}(p,q,r)=g_{\mu\nu}[\s_{wcA} +\S^{(wcA)}(p,q,r)]
+\tilde \G_{\mu\nu}^{(wcA)}(p,q,r)\,,
\\&&
\G^{(vcc)}(p,q,r)=-1+\s_{vcc}+ \S^{(vcc)}(p,q,r)\,,
\eeeqn
where $t_{\mu\nu}(p)\equiv p^2\,g_{\mu\nu}-p_\mu p_\nu$,
$$
t^{a_1 \cdots a_4}_{1;\mu_1 \cdots \mu_4}
=\left(\eps^{a_1a_2c}\eps^{ca_3a_4}-\eps^{a_1a_4c}\eps^{ca_2a_3}
\right)g_{\mu_1\mu_3}g_{\mu_2\mu_4} + (2\leftrightarrow 3) +
(3\leftrightarrow 4)
$$
is the four vector $SU(2)$ structure appearing in the BRS
action and
$$
t^{abcd}_{2;\mu\nu\r\s}=
(\de^{ab}\de^{cd}\,+\,\de^{ac}\de^{bd}\,+\,\de^{ad}\de^{bc})\;
(g_{\mu\nu} g_{\r\s}\, +\, g_{\mu\r} g_{\nu\s}\,+\,g_{\mu\s}
g_{\nu\r})\,.
$$
The relevant couplings are defined by the conditions
$$
\Sigma_{L}(0)=0\,,
\;\;\;\;\;\;\;\;\;
\frac{\partial \Sigma_{L}(p)}{\partial p^2}|_{p^2=\mu^2}=0\,,
\;\;\;\;\;\;\;\;\;
\Sigma_{T}(p)|_{p^2=\mu^2}=0\,,
$$
$$
\S^{(3A)}(p,q,r)|_{3SP}=0 \,,
\;\;\;\;\;\;\;\;\;
\S_1^{(4A)}(p,q,k,h)|_{4SP}=0\,,
\;\;\;\;\;\;\;\;\;
\S_2^{(4A)}(p,q,k,h)|_{4SP}=0\,,
$$
$$
\S^{(wc)}(p)|_{p^2=\mu^2}=0\,,
\;\;\;\;\;\;\;\;\;
\S^{(wcA)}(p,q,r)|_{3SP}=0 \,,
\;\;\;\;\;\;\;\;\;
\S^{(vcc)}(p,q,r)|_{3SP}=0\,.
$$
{}From these conditions\footnote{
Notice that in order to follow the general rule of extracting the
relevant part of a functional given by the operator $T_4^{(\mu)}$,
the couplings $\s_A$ and $\s_{wc}$ are defined in a slightly different
way with respect to ref.~\cite{YM}.
}
we can factorize in the vertices $\Sigma_i$
a dimensional function of $p$. Thus $\Sigma_i$ are ``irrelevant''
and contribute to the irrelevant part of the functional $\G[\phi,\g]$.
Similarly the vertices $\tilde \G_i$ are irrelevant since their
Lorentz structure is (partially in the case of $\tilde\G^{(4A)}$)
given by external momenta.

We recall that the ghost propagator and the $\bc$-$c$-$A$ vertex
are given in terms of the vertices
$\G_\mu^{(wc)}$ and $\G_{\mu\nu}^{(wcA)}$ by
$$
\G^{(\bc c)}(p)=
p^2 + i p^2 [\s_{wc}+\S^{(wc)}(p)]\,,
\;\;\;\;\;\;\;\;\;\;\;\;\;\;\;\;
\G_{\nu}^{(\bc cA)}(p,q,r)=-igp_\mu\,
\, \G_{\mu\nu}^{(wcA)}(p,q,r)\,.
$$

In subsect.~2.1  in order to better identify the physical
couplings in the effective action and study the gauge symmetry
we have made a different definition of the relevant part of the effective
action (see eq.~\re{Grel}), introducing the couplings $z_i$ and $\r_i$.
The relation of these couplings with the $\s_i$ is the following
$$
z_1=1-\s_A-\s_\alpha\,,\;\;\;\;\;
z_2=1+i\s_{wc}\,,\;\;\;\;\;
z_3=2-\frac i g \s_{3A}-\s_A-\s_\alpha\,,
$$
$$
\r_1=\s_{m_A}\,,\;\;\;\;\;
\r_2=\s_{\alpha}\,,\;\;\;\;\;
\r_3=1+i\s_{wc}+\s_{wcA}\,,
$$
$$
\r_4=\s_{vcc}+ i \s_{wc}\,,\;\;\;\;\;
\r_5=1-\s_{A}-\s_\alpha +\frac{1}{g^2} (\s_{4A}+\half \s'_{4A}) \,,
\;\;\;\;\;
\r_6=\frac {3}{2g^2} {\s'}_{4A}\,.
$$
In particular when the physical boundary conditions $z_1=z_2=z_3=1$
are imposed one finds
$$
\s_{\alpha}+\s_A=0\,,\;\;\;\;\;
\s_{wc}=0\,,\;\;\;\;\;
\s_{3A}= -ig \,,
$$
$$
\s_{m_A}=\r_1 \,,\;\;\;\;\;
\s_{\alpha}=\r_2 \,,\;\;\;\;\;
\s_{wcA}=-1+\r_3 \,,
$$
$$
\s_{vcc}= \r_4\,,\;\;\;\;\;
\s_{4A}= g^2 (-1+\r_5-\frac 1 3  \r_6)\,,\;\;\;\;\;
{\s'}_{4A}=\frac 2 3 g^2 \r_6 \,,
$$
which show that these boundary conditions fix the vector and ghost
wave function normalization and the three vector coupling.
The other couplings in $\G[\phi,\g]$ are given in terms of the $\r_i$
and determined by the symmetry as in section 4.

\vskip 1 true cm

\noindent
{\large\bf Appendix C}
\vskip 0.3 true cm
\noindent
We now extract the relevant part of the
most general one dimensional functional of fields and sources
with ghost number one. We call this generic functional $\D$. First of
all we isolate the vertices of $\D$ which contain the relevant couplings
\beeqn
\D &=& \int_p \D_\mu^{(Ac)}(p) A^a_\mu(-p) c^a(p)+
\frac 1 2 \eps^{abc} \int_p \int_q  \D_{\mu\nu}^{(AAc)}(p,q,r)
A^a_\mu(p) A^b_\nu(q) c^c(r) \\ &&
+ \frac 1 2 \eps^{abc} \int_p \int_q  \D_{\mu}^{(wcc)}(p,q,r)
w^a_\mu(p) c^b(q) c^c(r) \\ &&
+ \frac 1 2 \int_p \int_q \int_k
\D_{\mu\nu}^{(wAcc)abcd}(p,q,k,h)
w^a_\mu(p) A^b_\nu(q) c^c(k) c^d(h)\\ &&
+ \frac 1 6 \int_p \int_q \int_k
\D_{\mu\nu\r}^{(3Ac)abcd}(p,q,k,h)
A^a_\mu(p) A^b_\nu(q) A^c_\r(k) c^d(h)
+\ldots \,,
\eeeqn
where $r=-p-q$, $\,h=-p-q-k$ and the dots stand for the remaining terms
which are all irrelevant.
Then we define the $11$ relevant parameters as follows
\beeqn
\D_\mu^{(Ac)}(p)&=&p_\mu [\de_1+ p^2 \de_2+\D^{(Ac)}(p)]\,,
\\
\D_{\mu\nu}^{(AAc)}(p,q,r)&=&
g_{\mu\nu}(p^2-q^2)[\de_3+\D_1^{(AAc)}(p,q,r)] \\
&+& (p_\mu p_\nu - q_\mu q_\nu)[\de_{4}+\D_2^{(AAc)}(p,q,r)]
+\tilde{\D}_{\mu\nu}^{(AAc)}(p,q,r)\,,
\\
\D_{\mu}^{(wcc)}(p,q,r)&=&p_\mu[\de_5+\D^{(wcc)}(p,q,r)]\,,
\\
\D_{\mu\nu}^{(wAcc)abcd}(p,q,k,h)&=&g_{\mu\nu}
(\de^{ac}\de^{bd}-\de^{ad}\de^{bc})[\de_{6}+\D^{(wAcc)}(p,q,k,h)]
+ \tilde{\D}^{(wAcc)abcd}_{\mu\nu}(p,q,k,h)\,,
\\
\D_{\mu\nu\r}^{(3Ac)abcd}(p,q,k,h) &=& (\de^{ab}\de^{cd}p_\mu
g_{\nu\r}+\ldots)[\de_7+\D_1^{(3Ac)}(p,q,k,h)] \\
&+& (\de^{ad}\de^{bc}p_\mu g_{\nu\r}+\ldots)
[\de_8+\D_2^{(3Ac)}(p,q,k,h)] \\
&+& (\de^{ab}\de^{cd}p_\nu g_{\mu\r}+\ldots)
[\de_9+\D_3^{(3Ac)}(p,q,k,h)] \\
&+& (\de^{ac}\de^{bd}p_\nu g_{\mu\r}+\ldots)
[\de_{10}+\D_4^{(3Ac)}(p,q,k,h)] \\
&+& (\de^{ad}\de^{bc}p_\nu g_{\mu\r}+\ldots)
[\de_{11}+\D_5^{(3Ac)}(p,q,k,h)]
+\tilde{\D}^{(3Ac)abcd}_{\mu\nu\r}(p,q,k,h)\,,
\eeeqn
where the dots stand for permutation over the gluon momenta and
Lorentz and colour indices. The conditions defining the relevant
parameters are
$$
\D^{(Ac)}(0)=\partder{p^2}\D^{(Ac)}(p)|_{p^2=\mu^2}=0\,,
\;\;\;\;\;\;\;\;\;
\D_{1,2}^{(AAc)}|_{3SP}=0\,,
$$
$$
\D^{(wcc)}|_{3SP}=0\,,
\;\;\;\;\;\;\;\;\;
\D_{i}^{(3Ac)}|_{4SP}=0\,,
\;\;\;\;\;\;\;\;\;
\D^{(wAcc)}|_{4SP}=0\,.
$$
Due these conditions
one can isolate in these vertices a dimensional function of $p$, thus they are
irrelevant. Similarly the vertices ${\tilde \D}_i$ have the
Lorentz indices carried by momenta in a different way with respect to
their relevant parts and are irrelevant.

\vskip 1 true cm

\noindent
{\large\bf Appendix D}
\vskip 0.3 true cm
\noindent
In this appendix we prove that the fine tuning equations \re{ft'} can be
solved if $\DGr$ satisfies
\beq\nome{D0}
\DGr[\phi,\g;\delta_i]
=
{\cal S}_{{\G'}^{(0)}}\,
\Gr'[\phi,\g;\bar \r_i]
\,.
\eeq
An explicit calculation shows that eq.~\re{D0} implies the relations
\re{*2} and \re{*3} among the $\de_i$.
The parameters $\bar \r_i$ are given by the following
functions of $\delta_i$
\beq\nome{bard}
\eqalign{
\de_1=-\frac i g \br_1
\,,\;\;\;\;\;\;
\de_2&=-\frac i g \br_2
\,,\;\;\;\;\;\;
\de_3=\br_5
\,,
\cr
\de_5=-\frac{i}{g}(\br_6-\br_5)
\,,\;\;\;\;\;\;
\de_7&=ig (\br_5-\br_3+\br_4)
\,,\;\;\;\;\;\;
\de_8=-2ig (\br_5-\br_3)
\,.
}
\eeq
Now we show that it is possible to select
the six parameters $\r_i^{(\ell)}$ in
$\Gr[\phi,\g;\r_i^{(\ell)}]$
in such a way that six fine tuning equations, $\de_i^{(\ell)}=0$,
are satisfied.

{}From \re{ST} we have
$$
\D^{(\ell)}_{\G}[\phi,\g]
=2\,{\cal S}_{{\G'}^{(0)}}\,{\G'}^{(\ell)}
+\sum_{k=1}^{\ell-1} {\cal S}_{{\G'}^{(k)}}\,{\G'}^{(\ell-k)}\,.
$$
By applying $T^{(\mu)}_5$, we obtain the relevant part
$$
\Dr[\phi,\g;\de_i^{(\ell)}]
=
2\,{\cal S}_{{\G'}^{(0)}}\,{\Gr'}[\phi,\g;\r_i^{(\ell)}]
+\Omega^{(\ell)}[\phi,\g]\,,
$$
where
\beq\nome{Omega}
\Omega^{(\ell)}[\phi,\g]=T^{(\mu)}_5\,\sum_{k=1}^{\ell-1}
{\cal S}_{{\G'}^{(k)}}\,{\G'}^{(\ell-k)}
+2\, \left(
T^{(\mu)}_5\, {\cal S}_{{\G'}^{(0)}}
-{\cal S}_{{\G'}^{(0)}}\,T^{(\mu)}_4
\right)\;
{\G'}^{(\ell)}\,.
\eeq
The crucial observation now is that $\Omega^{(\ell)}$ does not depend on
the relevant parameters $\r_i^{(\ell)}$.
This is obvious, since the product of two relevant vertices is a
relevant vertex. This implies
$$
T^{(\mu)}_5\, {\cal S}_{{\G'}^{(0)}}\, T^{(\mu)}_4
= {\cal S}_{{\G'}^{(0)}}\, T^{(\mu)}_4
\,.
$$
As a consequence $(T^{(\mu)}_5\, {\cal S}_{{\G'}^{(0)}}
-{\cal S}_{{\G'}^{(0)}}\,T^{(\mu)}_4)\,{\Gr'}=0$ and
$\Omega^{(\ell)}$ does not receive contribution
from the couplings $\r_i^{(\ell)}$.
Thus the last term in \re{Omega} involves
only $\Gi^{(\ell)}[\phi,\g]$ which is given in terms of
$\r_i^{(\ell')}$ at lower loops $\ell'<\ell$.

Eq.~\re{D0} implies that $\Omega^{(\ell)}$ must be of the form
$$
\Omega^{(\ell)}=
{\cal S}_{{\G'}^{(0)}}\,
\Gr'[\phi,\g;{\r'}_i^{(\ell)}]\,,
\;\;\;\;\;\;
{\r'}_i^{(\ell)}=\bar \r_i^{(\ell)}-2\r_i^{(\ell)}
\,.
$$

As previously observed, ${\r'}_i^{(\ell)}$ are known from the
calculation of $\r_i^{(\ell')}$ at loop $\ell' < \ell$.
Therefore, by fixing
\beq\nome{sol}
\r_i^{(\ell)}=-\frac 1 2 {\r'}_i^{(\ell)}\,,
\eeq
we have $\bar \r_i=0$ and from \re{bard} we have the
final result
$\D^{(\ell)}_{\G}[\phi,\g]=0$.

\eject

\end{document}